# Interpretation of the modulus spectra of organic field-effect transistors with electrode overlap and peripheral regions: determination of the electronic properties of the gate insulator and organic semiconductor

Y. Suenaga,[1,a)] T. Nagase,[1,2] T. Kobayashi,[1,2] and H. Naito[1,2,b)]

[1]*Department of Physics and Electronics, Osaka Prefecture University, Sakai, Osaka 599-8531, Japan*

[2]*The Research Institute for Molecular Electronic Devices (RIMED), Osaka Prefecture University, Sakai, Osaka 599-8531, Japan*

The modulus spectra of organic field-effect transistors (OFETs) with electrode overlap and peripheral regions have been experimentally and theoretically investigated. The complex impedance of regioregular poly(3-hexylthiophene-2,5-diyl) (P3HT) OFETs with electrode overlap and peripheral regions was measured with a frequency response analyzer. The complex modulus was derived from an equivalent circuit of OFETs with overlap and peripheral regions using a four-terminal matrix approach. The modulus spectra of the P3HT OFETs were successfully fitted by those calculated using the expression derived from the equivalent circuit. Three structures were found in the modulus spectra of the P3HT OFETs owing to the dielectric properties of the gate insulator, transport properties of the organic semiconductor, and contact resistance from the low to high frequency ranges. The resistivity of the gate insulators and the field-effect mobility of working OFETs were determined using the values of the circuit components of the equivalent circuit obtained by fitting.

## I. INTRODUCTION

Organic field-effect transistors (OFETs) are currently attracting a great deal of attention as next-generation electronic devices because they can be fabricated by a printing process on flexible large-area substrates with low cost.[1–3] The device performance of OFETs, which has been usually investigated by analysis of the transfer and output characteristics of OFETs,[4,5] is closely related to the device structures. For instance, a staggered configuration of OFETs has often been used.[6–10] In this configuration, a certain overlap region between the source and the gate electrodes is essential to reduce the contact resistance, which is caused by the strong electric field formed between the source and the gate electrodes.[5] However, excessive electrode overlap regions can act as parasitic capacitance, and thereby the cut-off frequency of OFETs decreases with increasing area of the electrode overlap region.[11–13]

---

a) Electronic mail: yu.suenaga.oe@pe.osakafu-u.ac.jp.

Another origin of parasitic capacitance in OFETs is the peripheral region. The peripheral region of OFETs is defined as the region where an organic semiconductor is not covered by the source and drain electrodes, but by the gate electrode through the gate insulator.[14] It is not easy to obtain precise alignment between the organic semiconductor and the electrodes by printing processes, and thus to remove the peripheral region. Formation of a peripheral region is inevitable for OFET fabrication with a large process margin. In general, the peripheral region does not contribute to direct current (dc) operation in OFETs, but the greatly affects alternating current (ac) operation (reduction in the cut-off frequency with increasing area of the peripheral region).[12] The ac operation characteristics of OFETs without overlap and a peripheral region and with a peripheral region have been reported.[15–19] However, the influence of the overlap region on the OFET characteristics has not been explicitly investigated.[15–19]

It is therefore important to investigate the influences of both overlap and peripheral regions on the ac characteristics of OFETs. However, the influences of both peripheral and overlap regions on the complex impedance of OFETs have not been quantitatively investigated. A quantitative understanding of the complex impedance of OFETs is essential to successfully design OFET structures. Unlike the interpretation of the complex impedance of organic diodes, such as organic light-emitting diodes and organic photovoltaic cells,[20–22] the complex impedance of OFETs has been generally investigated by equivalent-circuit approaches.[23]

In this paper, we report the modulus spectra of OFETs with overlap and peripheral regions, and the modulus spectra are interpreted using an equivalent circuit of OFETs with overlap and peripheral regions. Interpretation of the modulus spectra is essential for understanding the ac operation characteristics of OFETs. The complex modulus was calculated from the equivalent circuit with transmission lines using a four-terminal matrix approach. The structures in the modulus spectra of the OFETs were determined and they were successfully reproduced by the calculated modulus spectra, from which the dielectric properties of the gate insulator, field-effect mobility of the organic semiconductor, and contact resistance were obtained. The influences of the overlap and peripheral regions on the structures in the modulus spectra were also investigated on the basis of the equivalent circuit.

## II. EQUIVALENT CIRCUIT WITH OVERLAP AND PERIPHERAL REGIONS

A small-signal equivalent circuit with electrode overlap that we developed is shown in Fig. 1(a). In Fig. 1(a), $L_c$ is the channel length, $L_o$ is the overlap length, $c_i$ is the dielectric capacitance per unit length, $r_s$ is the resistance of the semiconductor per unit length, $r_i$ is the resistance of the gate insulator per unit length, and $R_c$ is the contact resistance. For convenience, we define the impedance of the horizontal direction per unit length $z$ ($\Omega$ cm$^{-1}$) and the admittance of the vertical direction per unit length $y$ (S cm$^{-1}$) as follows:

$$z = r_s, \qquad (1)$$

$$y = \frac{1}{r_i} + j\omega c_i, \qquad (2)$$

where $j$ is the imaginary unit and $\omega$ is the angular frequency. First, we derive the complex modulus of the equivalent circuit with electrode overlap shown in Fig. 1(a) using a four-matrix approach.[24] The four-terminal matrix of circuits with a source–gate input terminal pair and a drain–gate output terminal pair is expressed as the cascaded connection of the channel and overlap regions:

$$[F]_{\text{total}} = [F]_{\text{overlap}}[F]_{\text{channel}}[F]_{\text{overlap}}. \qquad (3)$$

The four-terminal matrix of overlap region is

$$[F]_{\text{overlap}} = \begin{bmatrix} 1 + R_c y_0 & R_c \\ y_0 & 1 \end{bmatrix}, \qquad (4)$$

where $y_0 = L_0 y/2$ is the admittance of the overlap region. The four-terminal matrix of the channel region is[25]

$$[F]_{\text{channel}} = \begin{bmatrix} \cosh(L_c\sqrt{zy}) & \sqrt{z/y}\sinh(L_c\sqrt{zy}) \\ \sqrt{y/z}\sinh(L_c\sqrt{zy}) & \cosh(L_c\sqrt{zy}) \end{bmatrix}. \qquad (5)$$

$[F]_{\text{total}}$ is calculated from Eqs. (3)–(5). The distributed circuit of Fig. 1(a) can be converted to the lumped circuit of Fig. 1(b) by solving the following matrix equation:

$$[F]_{\text{total}} = \begin{bmatrix} \dfrac{z_1 + z_3}{z_3} & \dfrac{z_1 z_2 + z_2 z_3 + z_3 z_1}{z_3} \\ \dfrac{1}{z_3} & \dfrac{z_2 + z_3}{z_3} \end{bmatrix}. \qquad (6)$$

The modulus of the OFET whose drain and source electrodes are short-circuited is then calculated as the modulus of Fig. 1(b).
$M_{\text{total}} = j\omega(z_1/2 + z_3)$

$$= j\omega \cdot \frac{(1 + 3R_c y_0 + R_c^2 y_0^2)\cosh(L_c\sqrt{zy}) + \left(y_0\sqrt{\dfrac{z}{y}} + R_c\sqrt{\dfrac{y}{z}}\right)(1 + R_c y_0)\sinh(L_c\sqrt{zy}) + 1}{2y_0(2 + R_c y_0)\cosh(L_c\sqrt{zy}) + 2\left\{y_0^2\sqrt{\dfrac{z}{y}} + (1 + R_c y_0)\sqrt{\dfrac{y}{z}}\right\}\sinh(L_c\sqrt{zy})}. \qquad (7)$$

Next, we consider both the overlap and peripheral regions and derive the modulus expression. The equivalent circuit with overlap and peripheral regions is shown in Fig. 1(c). The contribution of the peripheral region can be regarded as admittance components connected in parallel to the overlap regions. In the following, we calculate the admittance of the peripheral regions. As shown in Fig. 1(d), we define the boundary between the overlap and peripheral regions as the input terminal and the edge of the gate electrode as the output terminal. The current and voltage at the input and output terminals in the peripheral region are then

$$\begin{bmatrix} V_{in} \\ I_{in} \end{bmatrix} = \begin{bmatrix} \cosh(L_p\sqrt{zy}/2) & \sqrt{z/y}\sinh(L_p\sqrt{zy}/2) \\ \sqrt{y/z}\sinh(L_p\sqrt{zy}/2) & \cosh(L_p\sqrt{zy}/2) \end{bmatrix} \begin{bmatrix} V_{out} \\ I_{out} \end{bmatrix}, \quad (8)$$

where $V_{in}$ and $I_{in}$ are the voltage and current at the input terminal, $V_{out}$ and $I_{out}$ are those at the output terminal, and $L_p$ is the length of the peripheral region. Because the output terminal in Fig. 1(d) is open, the admittance of the peripheral region $y_p$ can be expressed as

$$y_p = \left.\frac{I_{in}}{V_{in}}\right|_{I_{out}=0} = \sqrt{z/y}\cdot\tanh\left(\frac{L_p\sqrt{zy}}{2}\right). \quad (9)$$

The total modulus of the peripheral regions can be obtained by substituting $y_0' = y_0 + y_p$ as $y_0$ in Eq. (7):

$$M_{total} =$$

$$j\omega \cdot \frac{(1+3R_c y_0' + R_c^2 y_0'^2)\cosh(L_c\sqrt{zy}) + \left(y_0'\sqrt{\frac{z}{y}} + R_c\sqrt{\frac{y}{z}}\right)(1+R_c y_0')\sinh(L_c\sqrt{zy}) + 1}{2y_0'(2+R_c y_0')\cosh(L_c\sqrt{zy}) + 2\left\{y_0'^2\sqrt{\frac{z}{y}} + (1+R_c y_0')\sqrt{\frac{y}{z}}\right\}\sinh(L_c\sqrt{zy})}. \quad (10)$$

$(y_0' = L_0 y/2 + \sqrt{z/y}\cdot\tanh(L_p\sqrt{zy}/2))$

Setting $L_0 = 0$ and $L_p = 0$ in Eq. (10) then gives

$$M_{total} = j\omega \cdot \frac{\cosh(L_c\sqrt{zy}) + R_c\sqrt{\frac{y}{z}}\sinh(L_c\sqrt{zy}) + 1}{2\sqrt{\frac{y}{z}}\sinh(L_c\sqrt{zy})}. \quad (11)$$

This equation expresses the total modulus of the OFET without an overlap or a peripheral region (Fig. 1(e)). The expressions for the modulus of OFETs whose equivalent circuits are shown in Fig. 1(a), (c), and (e) have not been reported.

The calculated modulus spectra of the three different equivalent circuits in Fig. 1(a), (c), and (e) (with an overlap region, with peripheral and overlap regions, and without an overlap or a peripheral region, respectively) in the frequency range from 0.1 to 10 MHz are shown in Fig. 2. The parameters used for the calculation were $L_c = 250$ μm, $L_o = 100$ μm, $L_p = 450$ μm, $r_s = 1.6 \times 10^{10}$ Ω cm$^{-1}$, $r_i = 2.5 \times 10^8$ Ω cm, $c_i = 5.6 \times 10^{-10}$ F cm$^{-1}$, and $R_c = 7.0 \times 10^4$ Ω. The values are reasonable for poly(3-hexylthiophene) (P3HT) and the fluoropolymer CYTOP.

Unlike the equivalent circuit without overlap and peripheral regions, there are three structures in the imaginary part of the modulus spectra for equivalent circuits with an overlap region and with overlap and peripheral regions. Numerical analysis shows that the two peaks in the low and high frequency regions are because of the dielectric properties of the gate insulator and transport properties of the organic semiconductor, respectively. We can obtain the equivalent circuit components, such as $r_s$ and $c_i$, by fitting Eq. (10) to experimentally obtained modulus spectra of OFETs (the values of the geometrical parameters $L_c$, $L_o$ and $L_p$ are determined by optical microscope observation).

In Fig. 2(b), Re($M$) of OFETs with peripheral and overlap regions is expressed as

$$M_{\mathrm{LF}} = \frac{j\omega}{y \cdot (L_c + L_o + L_p)} \quad (12)$$

in the low frequency range from 1 to 100 Hz. The Re($M$) values of OFETs without peripheral and overlap regions and with an overlap region are expressed as

$$M_{\mathrm{LF}} = \frac{j\omega}{y \cdot L_c} \quad (13)$$

and

$$M_{\mathrm{LF}} = \frac{j\omega}{y \cdot (L_c + L_o)}, \quad (14)$$

respectively.

The calculated conductance and capacitance spectra of the three different equivalent circuits are shown in Fig. 2(c) and (d). In the conductance spectrum of the equivalent circuit with an overlap region, a peak is observed above $10^6$ Hz. However, there is no peak in this frequency range for the equivalent circuit without overlap and peripheral regions because the peak owing to the contact resistance is covered by the structure associated with the peak at 270 Hz owing to the transport properties of the organic semiconductor in the equivalent circuit without overlap and peripheral regions. The overlap region is hence essential to observe the structure owing to the contact resistance. The conductance spectra of the equivalent circuits without peripheral and overlap regions and with an overlap region are indistinguishable below $10^4$ Hz. However, there is a clear difference among the Im($M$) spectra for the three equivalent circuits. Because the contributions of the OFET material properties to the modulus spectra are clearly observed, we investigated the modulus spectra of the OFETs in the present study.

The complex impedance calculated from the equivalent circuit of the OFET without overlap and peripheral regions[15] can be well fitted to the capacitance and conductance spectra measured in OFETs with an overlap region.[17] The numerical investigation reveals that we can clearly distinguish the complex impedance of OFETs without overlap and peripheral regions and with an overlap region in the modulus plots.

## III. EXPERIMENT

A staggered OFET structure (Fig. 3), which is also referred to as the top-gate/bottom-contact (TGBC) configuration, was used for our OFET to reduce the contact resistance and hence improve the device characteristics. The channel length, overlap length, peripheral length, and channel width were $L_c$ = 250 μm, $L_o$ = 100 μm, $L_p$ = 350 μm, and $W$ = 1.5 mm, respectively.

Cr/Au source–drain electrodes were fabricated by photolithography and the lift-off process. Cr was vacuum deposited on a glass substrate as an adhesion layer, followed by Au evaporation. The thicknesses of Cr and Au were 3 and 40 nm, respectively. The P3HT layer was formed by spin coating a 1 wt% anhydrous chlorobenzene solution of regioregular P3HT ($M_w$ = 50000–100000, Sigma-Aldrich) at 2000 rpm. The P3HT thin film (thickness 40 nm) was dried in a Petri dish for 0.5 h and annealed in vacuum at 100 °C for 1 h. The P3HT thin film was prepared in a $N_2$-filled glove box. An insulating layer of a fluoropolymer, CYTOP (CTL-809M, AGC), was then spin-coated on the P3HT thin film as the gate insulator (thickness 500 nm). The CYTOP layer was dried in vacuum at 120 °C for 1 h. The CYTOP layer was formed using an orthogonal solvent, the CT-SOLV180 fluorine-based solvent (AGC). Finally, an Al gate electrode was evaporated on the CYTOP layer. The OFET fabricated in this way was annealed in vacuum at 60 °C for 1 day before the measurements for de-doping of oxygen and water in the devices.[26,27]

The impedance spectroscopy measurements were performed with a frequency response analyzer (Solartron ModuLab XM) in the frequency sweep range from $10^{-1}$ to $10^6$ Hz and in the gate bias voltage sweep range from −10 to −30 V. The source and drain electrodes were short-circuited, and the complex impedance between the gate electrode and the short-circuited source and drain electrodes was measured with the analyzer. The transfer characteristics were measured in the linear region of the OFETs using source meters (Keithley 6430 and 2400). All of the measurements were performed in vacuum at room temperature.

## IV. RESULTS AND DISCUSSION

The transfer characteristics in the linear region and the output characteristics of the top-gate P3HT OFET are shown in Fig. 4(a) and (b), respectively. The transfer characteristics exhibit negligibly small hysteresis and an on/off ratio of $10^3$. The field-effect mobility of the top-gate P3HT OFET was $1.1 \times 10^{-2}$ cm² V$^{-1}$ s$^{-1}$ (the mobility was calculated from the gradient of the transfer characteristic in the gate voltage range from −20 to −30 V). This value is almost the same as the mobility of conventional P3HT OFETs in the literature.[28,29]

The complex modulus spectra of the P3HT OFET at different gate voltages at room temperature are shown in Fig. 5. The two peaks in Fig. 5(a) shifted to higher frequency with increasing absolute value of gate bias voltage, while Im($M$) above 100 kHz was not affected by the gate bias voltage. The modulus spectra of the P3HT OFET at $V_g = -30$ V (solid circles) and the spectra calculated using Eq. (10) (solid lines) are shown in Fig. 6 to demonstrate that the calculated spectra are well fitted to the experimentally obtained spectra below $10^5$ Hz (Re($M$)) and $10^4$ Hz (Im($M$)). The reason for the disagreement between the experimental and calculated results above $10^5$ Hz (Re($M$)) and $10^4$ Hz (Im($M$)) will be discussed later. The experimentally obtained spectra are not well fitted by Eqs. (7) and (10). In addition, the complex modulus of the equivalent circuit with a peripheral region in the literature cannot be fitted to the experimental data in Fig. 5 because the overlap region was not taken into account in the literature.

The geometrical parameters $L_c$, $L_o$, and $L_p$, the capacitance of the gate insulator per unit length $c_i$, and $R_c$ in Eq. (10) are known. The fitting parameters in Eq. (10) are hence $r_s$ and $r_i$. The contact resistance does not contribute to the modulus spectra up to 100 kHz, and the modulus spectra from $10^{-1}$ to $10^5$ Hz can be reproduced with only two fitting parameters: $r_s$ and $r_i$.

We can determine the field-effect mobility in OFETs using the values of the equivalent circuit components. The electric charge $Q$ induced by $V_G$ and the current $I$ in the channel region are expressed as

$$Q = c_i L_c (V_g - V_{th}), \quad (15)$$

$$I = enS\mu E, \quad (16)$$

where $e$ is the elementary charge, $n$ is the carrier concentration in the channel, $S$ is the cross-sectional region of the channel, $\mu$ is the field-effect mobility, and $E$ is the electric field between the source and drain electrodes. The expression for the field-effect mobility from Eqs. (15) and (16) and the relation $r_s = E/I$ is

$$\mu = \frac{1}{r_s c_i (V_g - V_{th})}. \quad (17)$$

The gate-bias-voltage dependence of the field-effect mobility calculated by Eq. (17) is shown in Fig. 7. In Fig. 7, we also show the gate-bias-voltage dependence of the field-effect mobility calculated from the linear transfer characteristics of the P3HT OFET using the following equation:

$$\mu_e = \frac{\partial I_d}{\partial V_g} \cdot \frac{1}{W c_i} \cdot \frac{1}{V_d}, \quad (18)$$

where $I_d$ is the drain current and $V_d$ is the drain voltage. The field-effect mobility calculated by Eq. (17) is slightly higher than that calculated from the linear transfer characteristics in Fig. 4 using Eq. (18). This is mainly because of the contact resistance of the P3HT OFET. The field-effect mobility calculated from the linear transfer characteristics is reduced with increasing contact resistance,[30] while the numerical calculation shows that the field-effect mobility calculated by Eq. (17) is not greatly affected by the contact resistance, because the contact resistance is much smaller than the impedance of the gate insulator.

The resistivity of the gate insulator at different gate voltages obtained from the fitting is shown in Fig. 8. The resistivity of CYTOP decreases with increasing absolute value of gate bias voltage, and it is about five orders of magnitude smaller than the dc resistivity of CYTOP.[31] This is because the resistivity of CYTOP is dependent on the measurement frequency, as shown in Fig. 8 where the resistivity decreases with increasing measurement frequency. Generally, the frequency and temperature dependence of the conductivity of insulators and semiconductors can be expressed as

$$\sigma(\omega, T) = A\omega^s \qquad (19)$$

where $\sigma(\omega, T)$ is called the ac conductivity,[32,33] and $A$ and $s$ are temperature-dependent parameters. The ac conductivity can be explained by charge hopping between the localized states in the materials. We measured the conductivity of CYTOP in an Al/CYTOP/Al sandwiched device structure and found that the device exhibited frequency-dependent conductivity. The ac conductivity of the CYTOP sandwiched device at 0.1 Hz was almost the same as the dc conductivity of CYTOP.

We used the contact resistance value of 70 kΩ, which was obtained by the transmission line method,[34] in fitting the modulus spectra calculated by Eq. (10) to the experimental data in Fig. 6. As mentioned above, there is disagreement between the experimental and calculated results above $10^5$ Hz (Re($M$)) and $10^4$ Hz (Im($M$)) (Fig. 6). The origin of the disagreement is the limitation of the frequency response analyzer, where the modulus value in the frequency range above $10^5$ Hz may not be reliable. A frequency response analyzer with higher frequency response is therefore necessary to investigate the agreement between the experimental and calculated results in this frequency range, in which the modulus is mainly governed by the contact resistance, or to precisely determine the contact resistance.

As we have shown above, information about the electronic properties (the mobility of the organic semiconductor, contact resistance, and dielectric properties of the gate insulator) and geometrical parameters ($L_c$, $L_o$, and $L_p$) can be obtained from the measurements of the complex modulus of working OFETs. This information is essential for understanding the performance of working OFETs and to design OFETs.

**V. CONCLUSIONS**

We have reported expressions for the modulus of OFETs without overlap and peripheral regions, with an overlap region, and with overlap and peripheral regions. The expressions were derived from equivalent circuits of the OFETs using a four-terminal matrix approach, and they showed that the influence of the overlap and peripheral regions is clearly observed in the modulus spectra. We fabricated P3HT OFETs with overlap and peripheral regions and measured the complex impedance of the OFETs from 0.1 Hz to 10 MHz. The complex modulus spectra calculated from the modulus expression were well fitted to the experimentally obtained modulus spectra of the P3HT OFETs, and the important properties of the OFETs, the resistivity of the gate insulator, field-effect mobility, and contact resistance, were in principle determined by the fitting of the modulus expression to the experimental data. We demonstrated determination of the resistivity of the gate insulator, CYTOP, and the field-effect mobility of P3HT. The resistivity of CYTOP was lower than the dc resistivity of CYTOP, which results from the ac conductivity of CYTOP. The field-effect mobility of P3HT was slightly higher than that determined from the transfer characteristics of the P3HT OFETs because the measurements of the field-effect mobility from the modulus spectra were not significantly affected by the contact resistance. Unfortunately, the contact resistance was not accurately determined because of the frequency limitation of the frequency response analyzer used in the present study. We believe that equivalent-circuit analysis using the modulus expressions is of fundamental importance to investigate the performance of working OFETs.

**FIGURES**

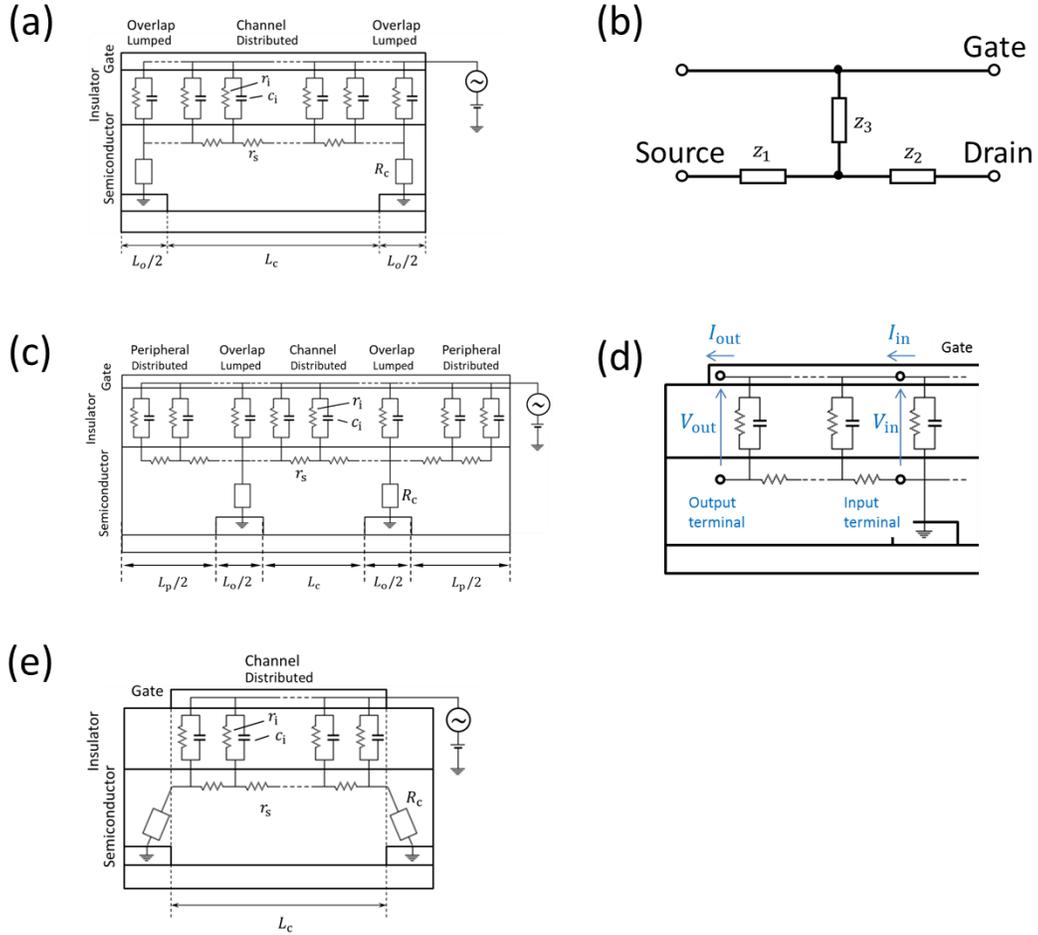

FIG. 1. (a) Equivalent circuit with an overlap region, (b) four-terminal lumped element circuit for derivation of Eq. (7), (c) equivalent circuit with overlap and peripheral regions, (d) equivalent circuit in the peripheral region, and (e) equivalent circuit without overlap and peripheral regions.

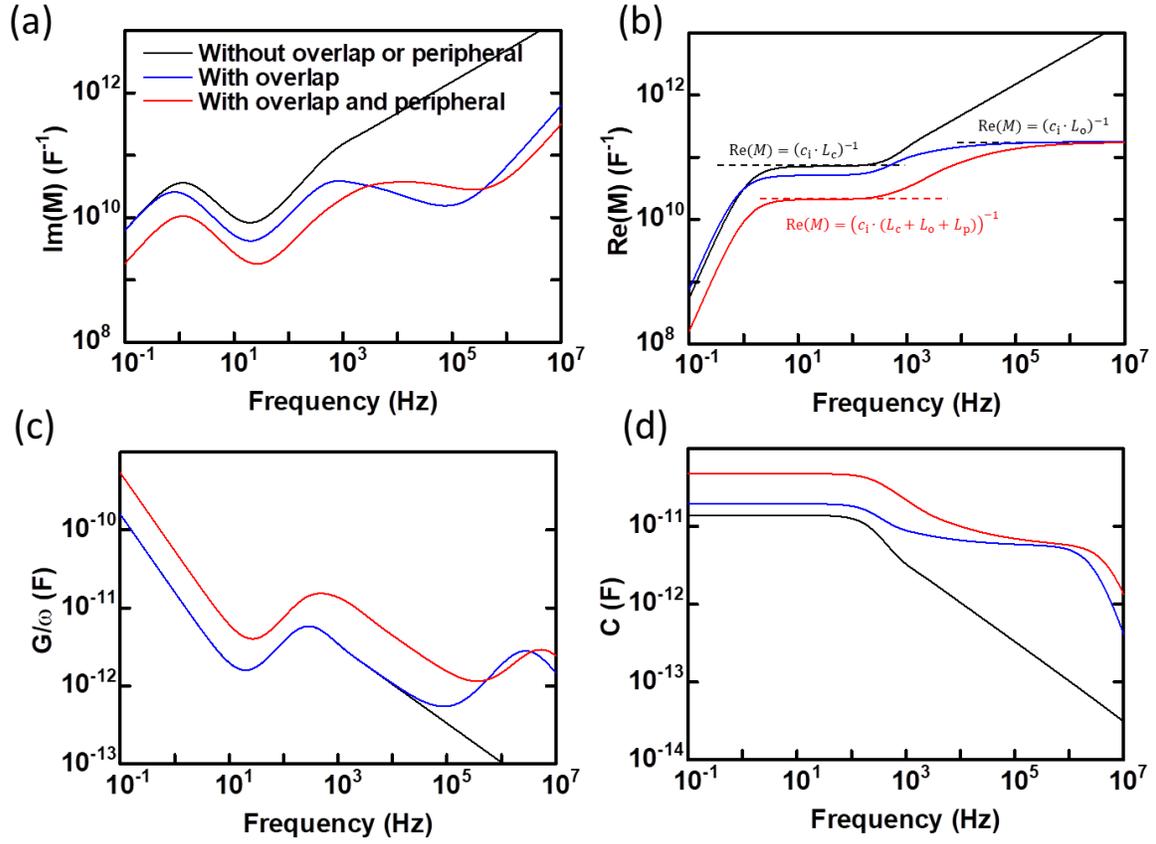

FIG. 2. Calculated (a) imaginary part of the complex modulus, (b) real part of the complex modulus, (c) conductance spectra, and (d) capacitance spectra for equivalent circuits without overlap and peripheral regions, with an overlap region, and with overlap and peripheral regions. The values of the circuit components used in the calculation were $L_c$= 250 µm, $L_o$= 100 µm, $L_p$= 450 µm, $r_s = 4.0 \times 10^9 \ \Omega \ cm^{-1}$, $r_i = 2.5 \times 10^8 \ \Omega \ cm$, $c_i = 5.6 \times 10^{-10} \ F \ cm^{-1}$, and $R_c = 7.0 \times 10^4 \ \Omega$.

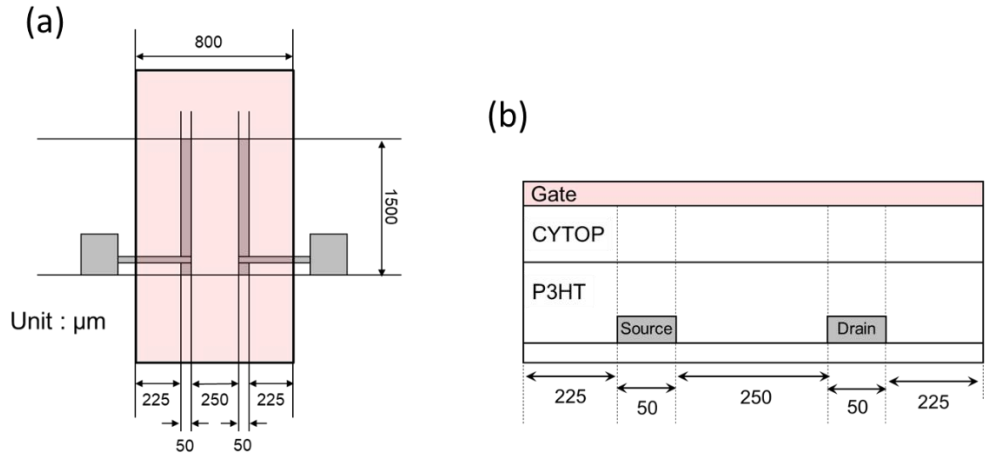

FIG. 3. (a) Top and (b) side views of the top-gate P3HT OFET.

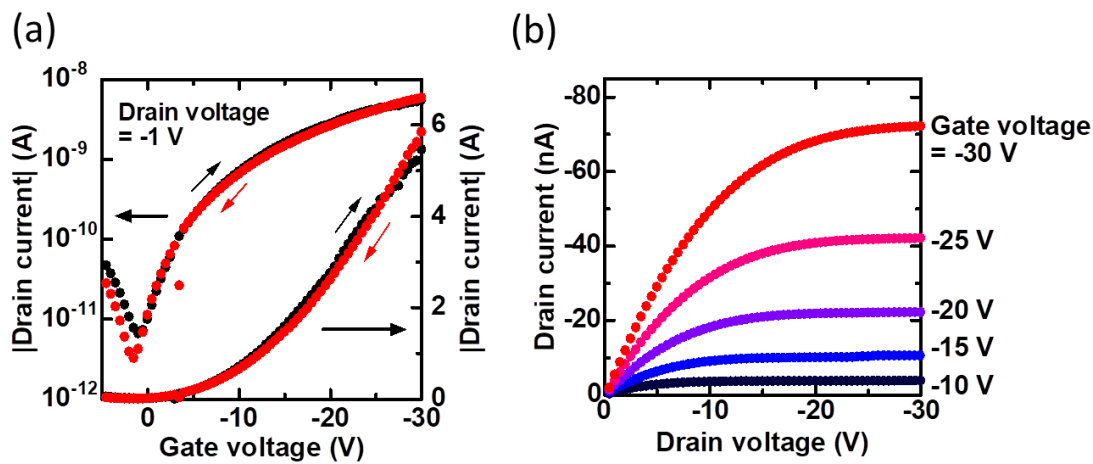

FIG. 4. (a) Transfer characteristics in the linear region and (b) output characteristics of the P3HT OFET at 297 K.

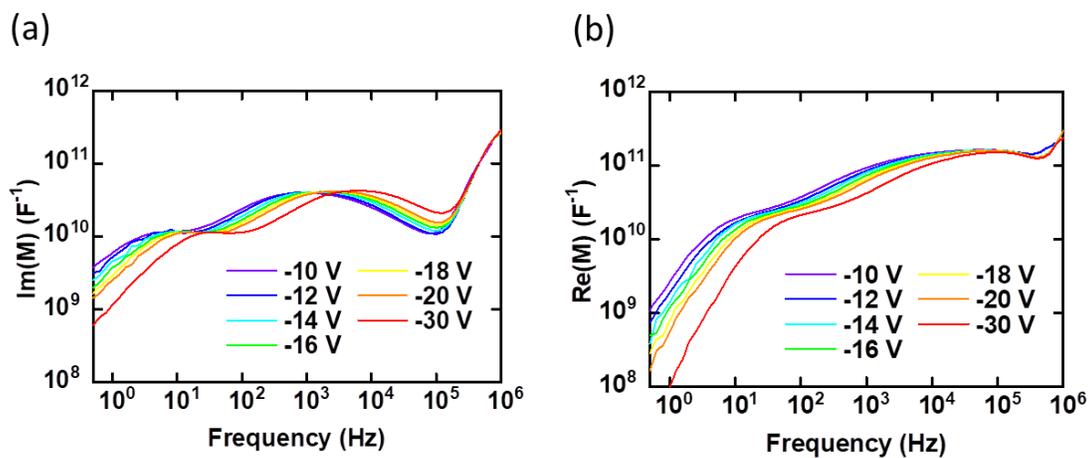

FIG. 5. Modulus spectra of the P3HT OFET at gate bias voltages from −10 to −30 V. (a) Imaginary and (b) real parts of the modulus spectra at 297 K.

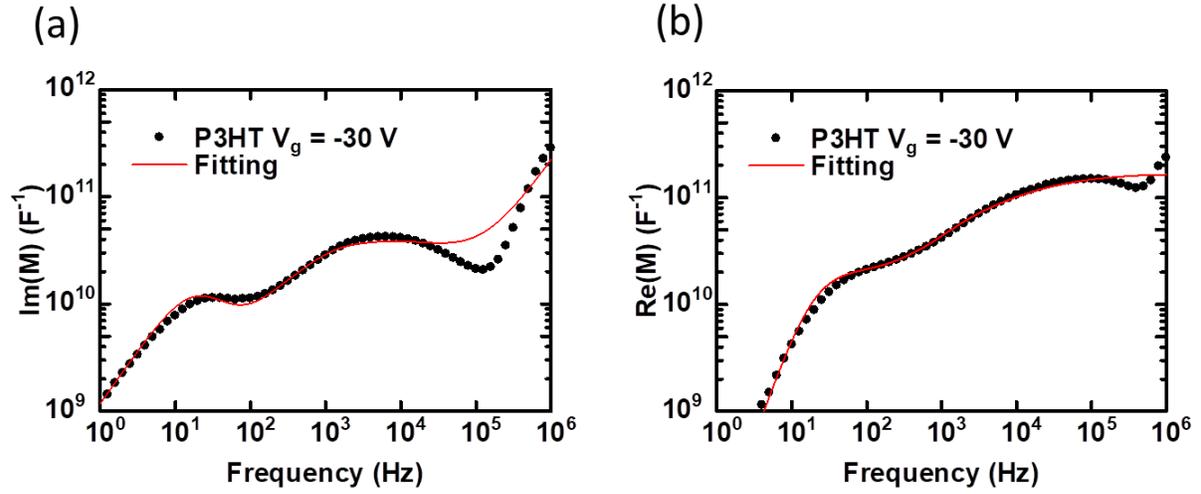

FIG. 6. Modulus spectra of the P3HT OFET at gate bias voltage = −30 V (black dots) and the fitted results by Eq. (10) (red lines). (a) Imaginary and (b) real parts of the modulus spectra at 297 K.

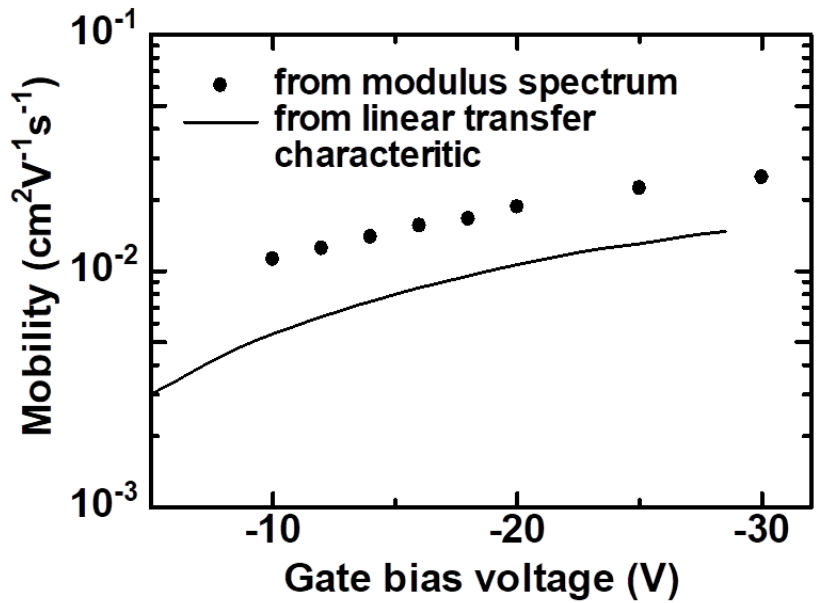

FIG. 7. Gate-bias-voltage dependence of the mobility calculated from the modulus spectra (black dots) in Fig. 5. The black line shows the mobility calculated from the linear transfer characteristics in Fig. 4.

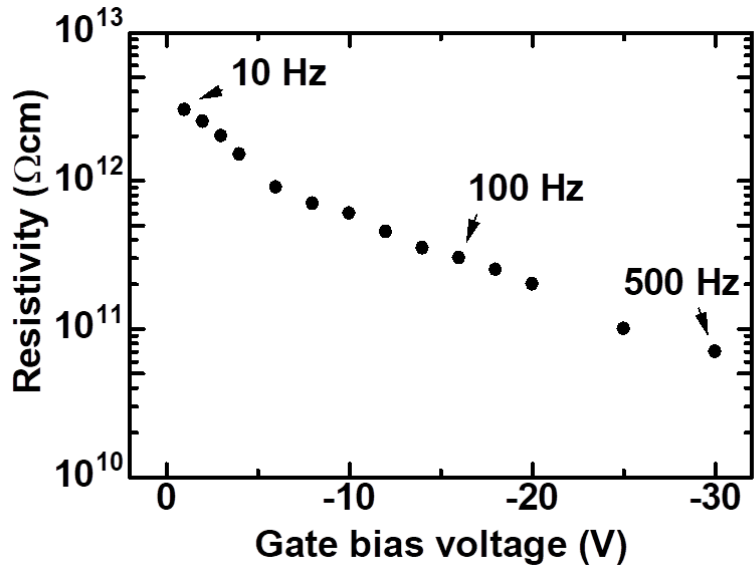

FIG. 8. Gate-bias-voltage dependence of the resistivity of the gate insulator determined by fitting Eq. (7) to the complex modulus spectra in Fig. 5. The frequencies shown in the figure are the peak frequencies owing to dielectric relaxation of the gate insulator, as shown in Fig. 5 (the peak shifted to higher frequency with increasing absolute value of the gate bias voltage).

**ACKNOWLEDGMENTS**

This work was partly supported by JSPS KAKENHI (Grant Nos. JP17H01265 and JP19H02599). We thank Tim Cooper, PhD, from Edanz Group (www.edanzediting.com/ac) for editing a draft of this manuscript.